%% file: main.tex
\documentclass[journal,10pt]{IEEEtran}
\usepackage{bbm}
\usepackage{amsmath}
\usepackage{acronym}  
\usepackage[dvips]{color}
\usepackage{epsf}
\usepackage{times}
\usepackage{epsfig}
\usepackage{notoccite}
\usepackage{graphicx}
\usepackage{epstopdf}
\usepackage{pstricks}
\usepackage{amssymb}
\usepackage{amsxtra}
\usepackage{here}
\usepackage{rawfonts}
\usepackage{float}
\usepackage{times}
\usepackage{url}
\usepackage{cite}
\usepackage{caption}
\usepackage{subcaption}
\usepackage{algorithm}
\usepackage{algpseudocode}
\usepackage{blindtext}
\usepackage{enumitem}
\usepackage{xcolor,cite,etoolbox}
\usepackage{relsize}
\usepackage{lipsum}
\usepackage{graphicx}
\usepackage{tabularx}
\usepackage{xparse}
\usepackage{array}
\newcolumntype{P}[1]{>{\centering\arraybackslash}p{#1}}
\usepackage{mleftright}

\usepackage{mathtools}

\usepackage{graphics}
\usepackage{physics}
\usepackage{amssymb}
\usepackage{siunitx}
\include{newcommands}
\usepackage{multicol}

\usepackage[nomain,acronym,shortcuts]{glossaries}
\makeglossaries
\newcommand*{\acro}[3][]{\newacronym[#1]{#2}{#2}{#3}}
\include{acronyms}
\usepackage{datetime}
\usepackage{amssymb}
\usepackage{subcaption}
\usepackage{verbatim}

\IEEEoverridecommandlockouts

\begin{document}
\title{The Seven Worlds and Experiences of the Wireless Metaverse: Challenges and Opportunities
\thanks{This research was supported by the U.S. National Science Foundation under Grants CNS-2210254 and CNS-2007635.}
\thanks{O. Hashash, C. Chaccour, and W. Saad are with Wireless@VT, Bradley Department
of Electrical and Computer Engineering, Virginia Tech, Arlington, VA, USA,
Emails: omarnh@vt.edu, christinac@vt.edu, walids@vt.edu.}
\thanks{T. Yu and K. Sakaguchi are with the Department of Electrical and Electronic Engineering, Tokyo Institute of Technology, Tokyo, Japan, Emails: yutao@mobile.ee.titech.ac.jp, sakaguchi@mobile.ee.titech.ac.jp.}
\thanks{M. Debbah is with the Technology Innovation Institute, 9639 Masdar City, Abu Dhabi, United Arab Emirates and also with CentraleSupelec, University Paris-Saclay, 91192 Gif-sur-Yvette, France, Email: merouane.debbah@tii.ae.}}%

\author{Omar~Hashash,~\IEEEmembership{Graduate~Student~Member,~IEEE,}
Christina~Chaccour,~\IEEEmembership{Member,~IEEE,} 
Walid~Saad,~\IEEEmembership{Fellow,~IEEE,}
Tao~Yu,~\IEEEmembership{Member,~IEEE,}
Kei~Sakaguchi,~\IEEEmembership{Senior~Member,~IEEE,}
and~M{\'e}rouane~Debbah,~\IEEEmembership{Fellow,~IEEE}\vspace{-0.8cm}}%
\maketitle

\begin{abstract}
The wireless metaverse will create diverse user \emph{experiences} at the intersection of the \emph{physical, digital, and virtual worlds}. These experiences will enable novel \emph{interactions} between the \emph{constituents} (e.g., extended reality (XR) users and avatars) of the three worlds. 
However, remarkably, to date, there is no holistic vision that identifies the full set of metaverse worlds, constituents, and experiences, and the implications of their associated interactions on next-generation communication and computing systems.
In this paper, we present a holistic vision of a \emph{limitless, wireless metaverse} that distills the metaverse into an intersection of \emph{seven worlds and experiences} that include the:~\emph{i) physical, digital, and virtual worlds}, along with the \emph{ii) cyber, extended, live, and parallel experiences}.
We then articulate how these experiences bring forth interactions between diverse metaverse constituents, namely, a) \emph{humans and avatars} and b) \emph{connected~intelligence~systems and their digital~twins~(DTs)}. 
Then, we explore the wireless, computing, and artificial intelligence (AI) challenges that must be addressed to establish \emph{metaverse-ready} networks that support these experiences and interactions. We particularly highlight the need for end-to-end synchronization of DTs, and the role of human-level AI and reasoning abilities for cognitive avatars.
Moreover, we articulate a sequel of open questions that should ignite the quest for the future metaverse. We conclude with a set of recommendations to deploy the limitless metaverse over future wireless systems. 
\vspace{-0.1cm}
\end{abstract}
\begin{IEEEkeywords}
metaverse, avatars, extended reality (XR), connected intelligence systems (CISs), digital twins (DTs)
\vspace{-0.5cm}
\end{IEEEkeywords}

\section{Introduction}
\vspace{-0.15cm}
The \emph{metaverse}, that sits at the cross-road of the \emph{physical, digital,} and \emph{virtual realms}, will enable a multitude of world \emph{experiences} that allow users to travel across space and time. Metaverse user experiences are realized at the intersections of the metaverse worlds thereby enabling many socially impactful applications.
For example, at the intersection of all three worlds, an \ac{XR} user can be seamlessly teleported with their senses to visit the world using a multi-sensory avatar.
Despite this promising potential of the metaverse, to date, the state-of-the-art~\cite{wang2022survey} restricts its application space to the individual physical, digital, and virtual worlds, without exploring the opportunities offered by their intertwined experiences.
Moreover, the current literature often neglects the role of key constituents that share the metaverse along with \ac{XR} users and avatars. For example, \acp{CIS} (e.g., autonomous vehicles, robots, etc.) actively intervene in the metaverse by acquiring \acp{DT} to proactively control their autonomous physical agents~\cite{khan2022digital}, a challenging aspect that is less understood in the metaverse literature.

Naturally, taking into account new experiences and constituents introduces unique challenges when deploying the metaverse in the real world. Chief among those challenges is the novel set of \emph{interactions} between the constituents of the different worlds created by the metaverse experiences. 
Those include two distinct types: 1) between \ac{XR} users and avatars and, 2) between \acp{CIS} and 
\acp{DT}. As a byproduct of such interactions, the metaverse should be extended beyond its traditional human-oriented boundaries to embrace other constituents residing in the real world (e.g., \acp{CIS}, physical assets, biological systems, etc).
Indeed, engineering a \emph{limitless metaverse} to cater for all types of experiences and interactions necessitates transitioning from \emph{human-centric} to \emph{everything-centric} designs. This, in turn, requires overcoming several unique wireless, computing, and \ac{AI} challenges. For example, the fact that avatars must be aware of their \ac{XR} users introduces new \ac{AI} challenges requiring human-level intelligence and cognition. Meanwhile, the need for real time configuration of \acp{CIS} via \acp{DT} imposes stringent \ac{E2E} synchronization requirements such as near-zero latency and ultra high data rates to teleport massive physical entities into the metaverse.

\begin{figure*}
	\centering
	\includegraphics[width=0.795\linewidth]{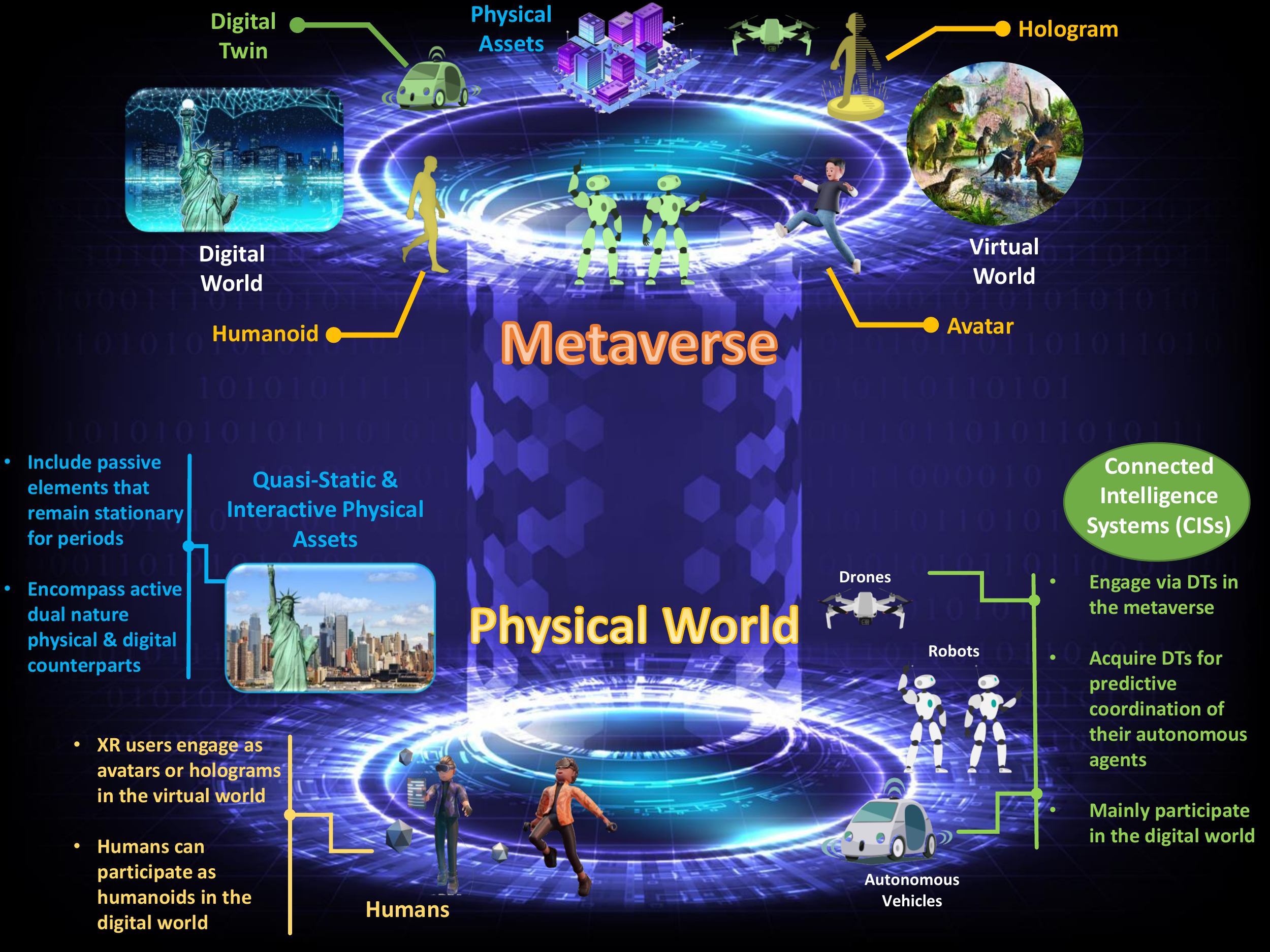}
	\caption{\small{Illustration of the limitless wireless metaverse comprising different worlds and diverse constituents.}}
	\label{Metaverse Vision}
 \vspace{-0.5cm}
\end{figure*}

Prior works~\cite{lam2022human,khan2023metaverse,wang2022wireless} attempted to investigate the interactions in the metaverse, but they are mainly limited to physical modeling techniques of humans as avatars (e.g.,~\cite{lam2022human} and \cite{khan2023metaverse}) and avatar construction methods~\cite{wang2022wireless}.
Hence, these works cannot capture the metaverse's \emph{human-to-avatar} interaction that defines the first stage in the interaction with other avatars and network elements.
Moreover, this prior art 
~\cite{lam2022human,khan2023metaverse,wang2022wireless}
does not design the digital world as a true replica of the real world. Instead, it conflates the virtual and digital paradigms. Consequently, these works do not fully capture the \emph{CIS-to-DT} interactions between the physical and digital worlds.
\emph{Evidently, there exists a gap in developing a rigorous metaverse vision that precisely recognizes its constituents, worlds, and experiences along with their implications on communication, computing, and \ac{AI} system designs.}


\indent The main contribution of this paper is to fill this gap by charting a \emph{holistic vision of a limitless, wireless metaverse}, which unlocks the full set of metaverse worlds and their constituents. We particularly articulate how this boundless vision can generate novel experiences between worlds, driving interactions between the constituents over the network.
Our contributions include:
\vspace{-0.12cm}
\begin{itemize}
    \item We transform the metaverse from a vague ensemble of worlds, as done in prior works, into an intersection of \emph{seven worlds and experiences}, namely, \emph{physical, digital, and virtual worlds}, with \emph{cyber, extended, live, and parallel experiences}~(see Fig.~\ref{Metaverse Experiences}), that have unique applications and challenges.
    \item From these experiences, we identify major challenges pertaining to the \emph{interactions} between the constituents of the different worlds. We rigorously define the \emph{human-to-avatar} and \emph{\ac{CIS}-to-\ac{DT}} interactions from the physical world into the virtual and digital worlds, respectively.
    \item We explore the wireless and computing challenges needed to create metaverse-ready networks that support the identified experiences and interactions. We delineate key challenges such as achieving \ac{E2E} synchronization between \ac{XR} users and avatars on the network, as well as providing ultra-synchronization high-rate low latency communications to support metaverse interactions.
    \item We identify key \ac{AI} challenges including the de-synchronization of \acp{DT}, the need for resilient designs that accounts for it, and the tradeoff between catastrophic and graceful forgetting for generalizable \acp{DT}.
    \item We articulate a set of open questions and challenges that motivate the quest towards a next-generation metaverse system, and finally conclude with key recommendations.
\end{itemize}
\vspace{-0.45cm}

\section{Metaverse Vision: Breaking Down the Worlds, Constituents, and Experiences}
\label{Vision}
To capture the peculiarity of the metaverse, we unfold its \emph{worlds}, and specify their corresponding \emph{constituents} (see Fig.~\ref{Metaverse Vision}), and delineate the \emph{experiences} arising at their intersection (see Fig.~\ref{Metaverse Experiences}).
\vspace{-0.5cm}

\subsection{The Metaverse Worlds}
\vspace{-0.05cm}
\subsubsection{Physical World} 
This world is a subset of the real world that encompasses its human, biological, machine, and system fabrics.

\subsubsection{Digital World} 
The extent of the digital world broadly entails a duplicate of the real world \emph{in the digital domain}. The digital world mirrors the physical world to enable an alternative \emph{digital reality}.
Hence, to acquire a high fidelity digital representation of the real world, real world elements are massively scaled with sensing abilities to confidently replicate their characteristics digitally. The duplicated elements constitute synchronous twin-like representations, however, \emph{in the soft sense of bits and bytes}.   

\subsubsection{Virtual World}
In compliance with the \emph{parallel universe theory}, a virtual world is a \emph{multiverse}, i.e., a collection of artificially synthesized hyperspaces, fabricated on a fictional plane of imaginary elements. This world is composed of \emph{visualizable} elements that resemble those of the real world, \emph{in shape}, however, they are purely synthetic, \emph{in nature}. Accordingly, one can conceive that this world includes the set of a) enhanced online platforms (e.g., Roblox), b) social working environments (e.g., Meta Horizon), and c) supplementary assets for enterprise services (e.g., lands, stores, etc.).
\vspace{-0.4cm}

\subsection{Real World Constituents and their Representations}
\vspace{-0.1cm}
Decomposing the metaverse into different worlds demands pinpointing the \emph{physical} world constituents and mapping them to their \emph{digital} and \emph{virtual} representations:
\subsubsection{Humans}
To engage in the metaverse, humans are represented in multiple forms depending on the world where they reside:
\begin{itemize}
    \item \textbf{Avatars:} 
    One common human surrogate in the virtual world is in the form of avatars. An avatar is basically a 3D human-like bot that allows users to interact and attain visualization from peers using \ac{XR} devices. 
    However, due to the real-time synchronization and embodiment of senses and feelings vital to preserve the human-to-avatar duality, avatars must inevitably use \emph{human-driven \ac{AI} or control systems}.

    \item \textbf{Humanoids:}
    We define \emph{humanoids} as \emph{massively sensed matterless humans}. Unlike avatars, who embody the \ac{XR} user in the virtual world, humanoids capture the human existence, in the digital world,  without their involvement. In other words, humanoids are digital representations of humans composed of a void nature. Hence, humanoids play a critical role in defining the mechanism of digital applications (e.g., \ac{DT} configuration mechanism).
    \item \textbf{Holograms:}
    Human holograms will be a key metaverse constituent. Holograms are 3D projected images synthesized by capturing the light reflected over human entities. Using holographic technology, humans can become omnipresent in the virtual world, once their impinged light beams are captured and reconstructed.
\end{itemize}

\subsubsection{\acp{CIS}}
In our envisioned metaverse, complex cyber-physical agents (e.g., autonomous vehicles, drones, robots, etc.) will inevitably share the real world with humans. \Acp{CIS} require \acp{DT} to proactively configure their autonomous physical agents and optimize their decision making process.
We emphasize that, in the metaverse, \emph{the role of \acp{DT} is far beyond a mere replication of physical world elements}. In contrast, \acp{DT} -- in the massive twinning context -- serve as sophisticated, \emph{large-scale, bidirectional operational \ac{AI} models} that can proactively control, predict, and configure the states of numerous autonomous systems~\cite{hashash2022edge}. These \acp{CIS} make decisions based on their interactions with the environment (e.g., through reinforcement learning or other means). Meanwhile, a \ac{CT} is simulated in the digital world to mimic the functionalities of its \ac{PT}. For providing real-time predictive coordination, advanced \ac{AI} tools are needed to forecast the future state of the \ac{CT}. Based on this predicted state, \ac{PT} configurations are proactively adjusted, enabling seamless blending between physical and digital worlds.

\subsubsection{Quasi-static and interactive physical assets}
The metaverse's physical world also includes \emph{physical assets}. 
Unlike \acp{CIS}, these assets cannot be replicated into the metaverse with the same level of sophistication. While \acp{DT} of \acp{CIS} utilize \ac{AI} models to actively control the state of autonomous systems, physical assets lack this interactivity which
limits their presence to \emph{unidirectional simulation streams}. Physical assets are  broadly grouped into two categories:
\begin{itemize}
    \item \textbf{Quasi-static physical assets:} These are representations of physical elements that remain \emph{stationary} for a periodic duration (e.g., buildings, statues, etc). In other words, their rate of change is minimal with respect the rate of change of other constituents. Generally, these elements require massive sensing abilities to be teleported into the metaverse. Quasi-static physical assets are also \emph{passive} (e.g., any change on the digital replica of the statue of liberty in the metaverse does not reflect back to the real world).
    \item \textbf{Interactive physical assets:} Interactive assets comprise the set of \emph{active} elements that have a \emph{dynamic} dual nature of physical and digital counterparts. These assets require a continuous, real-time information pipeline between the counterparts to keep them synchronized. A key example would an interactive machine in Industry $5.0$ that is physically present in the real world to be controlled via human intervention. Simultaneously, this machine is digitally teleported into the metaverse for remote interactions with \ac{XR} users. Here, any minute change from one counterpart is promptly reflected onto the other.
\end{itemize}

\begin{figure*}
	\centering
	\includegraphics[width=0.795\linewidth]{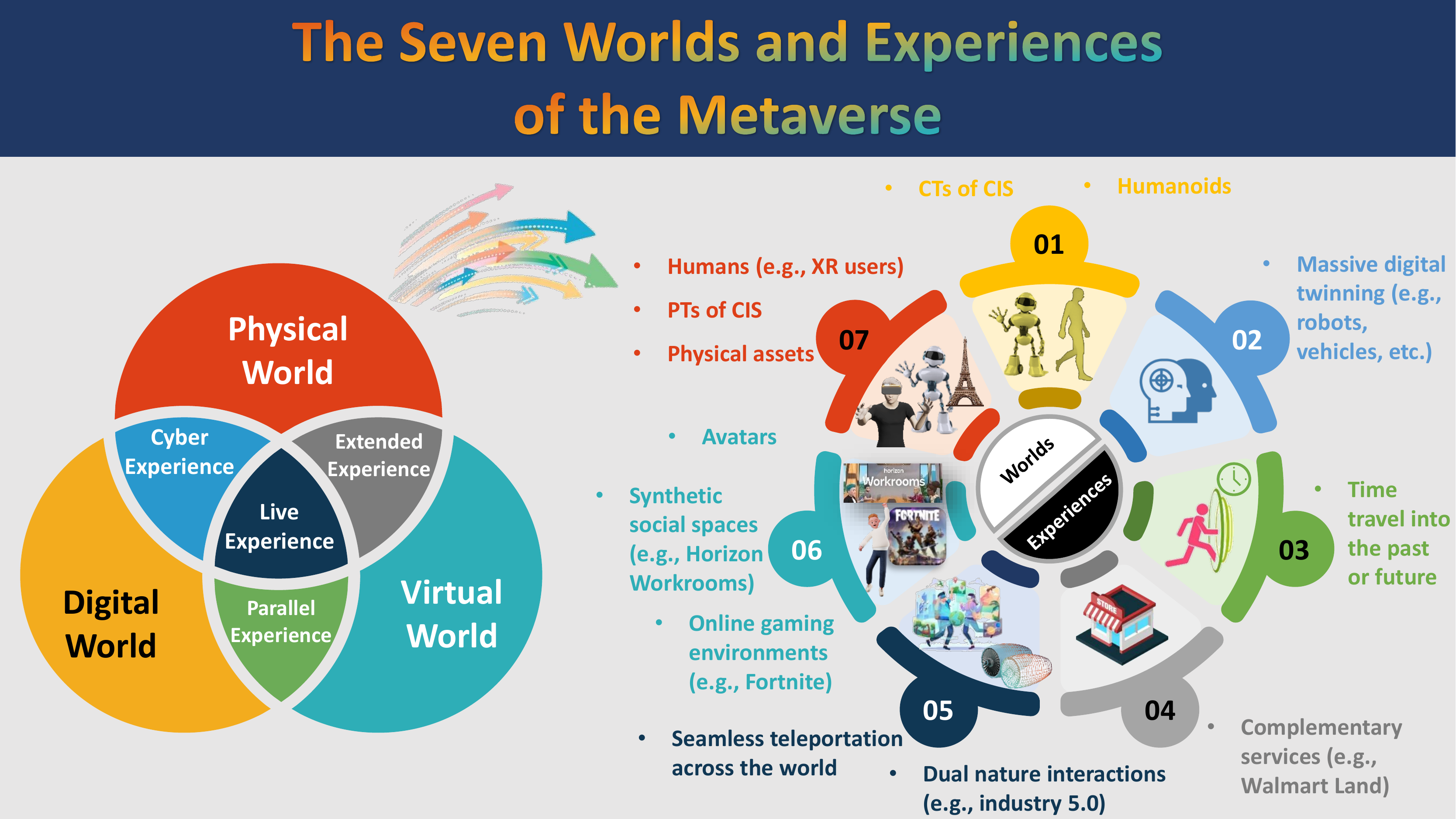}
	\caption{\small{Illustration of our envisioned seven worlds and experiences of the wireless metaverse.}}
	\label{Metaverse Experiences}
 \vspace{-0.48cm}
\end{figure*}
\vspace{-0.35cm}
\subsection{On the Experiences Between the Worlds}
\vspace{-0.05cm}
As the worlds collide to create a novel set of metaverse applications, different \emph{experiences} are realized at every intersection (see Fig.~\ref{Metaverse Experiences}): 
\subsubsection{Cyber Experience}
The overlap between physical and digital realms creates a new cyber experience to enable \acp{CIS} and cyber-physical systems that admit \acp{DT} throughout their life-cycle. In this experience, \acp{PT} and human interventions occur in the physical world, while the corresponding \acp{CT} reside in the digital world.

\subsubsection{Extended Experience}
In an extended experience, the physical and virtual boundaries interlapse to extend the real world, thereby providing supplementary virtual services or assets. Prime examples here lie within the industrial and enterprise metaverse. For example, enterprise stores (e.g., Walmart Land) can include merchandise attributed to the real world, however, they may be commercialized in a virtual store that extends the geospatial store from the physical world. This is also the ground for the bulk of state-of-art virtual applications that represent the \ac{XR} user as a virtual world avatar.  

\subsubsection{Live Experience}
The crossroad of all worlds represents a focal point for an online experience that recreates real life in different spatial settings.
Live experiences enable seamless holographic teleportation across different locations worldwide.
Another example of live experiences includes Industry $5.0$ applications. For instance, a physical machine can be duplicated into the digital world and visualized as a virtual element accessible to \ac{XR} users for intervention.

\subsubsection{Parallel Experience}
Upon merging the digital and virtual worlds, one can blend the massive real world data from the digital world history with generative \ac{AI} tools to witness a travel to parallel time zones within the bounds of the virtual world. In such parallel experiences, one can experience time travel to different spaces, into the past or future times, through an artificial warping of space-time dimensions.

\vspace{-0.26cm}
\section{Human-to-Avatar and \ac{CIS}-to-\ac{DT} Interactions in the Metaverse}
The experiences of Section~\ref{Vision} will drive a set of interactions between metaverse constituents, namely, human-to-avatar and \ac{CIS}-to-\ac{DT} interactions, described next.
\vspace{-0.4cm}
\subsection{Human-to-Avatar Interactions: Towards Cognitive Avatars} 
\vspace{-0.05cm}
Realizing the affinity between \ac{XR} users and avatars hinges on the human-to-avatar interactions that allow avatars to faithfully \emph{embody} their respective \ac{XR} users when interacting with peers in the virtual world. For instance, to maintain a robust duality, the avatar should promptly mimic its \ac{XR} user. This is carried out in two modes: 1) \emph{forward mode} from \ac{XR} user to avatar and 2) \emph{backward mode} from avatar to \ac{XR} user.   
The avatar should be able to announce this duality in emulating the \ac{XR} user actions (forward mode) and actuating the repellent feedback (backward mode). 
Embodying the \ac{XR} user in the avatar requires more than just \emph{blindly copying} the estimated user position and movements as in~\cite{lam2022human}. While this may partially suffice in the forward mode~\cite{khan2023metaverse}, it fails to account for the backward mode in which the feedback from the avatar to the \ac{XR} user should be \emph{in sync}. To address this challenge, the avatar must become \emph{cognizant} of the \ac{XR} user actions, by \emph{understanding the underlying logic stemming from the sensory inputs that initiated them.} Indeed, a \emph{cognitive avatar} should \emph{learn the persona of its underlying \ac{XR} user} and \emph{mimic human intelligence}. This is accompanied with the need to transfer the actions and feedback in between \ac{XR} user and avatar.

To perfectly embody the \ac{XR} user in the forward mode, it is vital to conserve the \emph{synchronization, accuracy, and precision} in duplicating the actions to the avatar. One way to tackle this duality is through a \emph{mirror game}~\cite{lombardi2021using} -- a powerful framework for investigating the social motor coordination between two human players. This concept can be extended by substituting one of the players with an \emph{\ac{AI}-driven cognitive avatar}. Accordingly, the players adhere to a \emph{leader-follower} strategy, where the avatar (follower) learns the unique \emph{kinematic fingerprint} that characterizes how the \ac{XR} user (leader) exhibits movements and actions in response to their sensory and tracking information (e.g., through imitation learning). The ultimate goal in this forward mirror game is acquiring an \ac{AI}-driven avatar that minimizes the mismatch in replicating the instantaneous \ac{XR} actions in terms of accuracy and synchronization.

In the backward mode, the feedback impinging from peer avatars (and virtual elements) should be synchronously reflected to the \ac{XR} user. This interaction is translated into \emph{senses} and \emph{actuations}. For example, if an avatar is punched on their arm, then this ``punched'' arm should move \emph{in sync} with that of the \ac{XR} user. 
To perform this process, the feedback to the \ac{XR} user is reflected via a \emph{reverse mirror game}. 
Flipping the roles, the avatar (leader) in this backward game will use its acquired knowledge, from the forward mode, to \emph{reason for and execute the feedback}, and further pass the corresponding senses and actuations to the \ac{XR} user (follower). Hence, the avatar requires \emph{abductive reasoning} capabilities to \emph{inversely} reach the senses and actuation inferences that the \ac{XR} user would most likely feel and experience due to the feedback. Therefore, the overall interaction should be modeled as a \emph{bidirectional mirror game} that integrates the synergies in the forward and backward games.

Embodiment requires not only overcoming the mismatch challenges, but also promoting technologies such as the \emph{Internet of Senses} and \emph{Internet of Feelings}. 
In fact, enabling somatosensation for cognitive avatars requires harmonizing the senses to achieve \emph{semantic congruence}, i.e., agreement between the meanings of senses.
Thus, synchronization (in time) of the senses at the \ac{XR} user level is necessary to provide the desired harmony. Moreover, an agreement between the \emph{meanings} of perceived senses and their effective overall stimuli is necessary to ensure the true reception of senses. Eventually, it is the harmonization between \emph{sensation} and \emph{perception} that plays a crucial role in embodying the avatar.
Moreover, cognitive avatars should be able to express the feelings of the \ac{XR} users in their interaction with peers. This relies on advances in \emph{affective computing} and \emph{emotion \ac{AI}}~\cite{bernal2017emotional}. Affective computing allows the avatar to reflect the physiological state of the \ac{XR} user -- a cornerstone for enabling viable avatar interactions.

\vspace{-0.35cm}
\subsection{\ac{CIS}-to-\ac{DT} Interaction: A Multi-View Generative \ac{AI} Approach }
\vspace{-0.2cm}
Another key metaverse interaction occurs among \acp{CIS} and DTs. The \ac{CIS}-to-\ac{DT} interaction involves the \ac{DT} prediction mechanism that proactively configures the (physical) autonomous \ac{CIS} agents in real time, as explained in Section~\ref{Vision}. However, predicting the future state of a \ac{CT} is a complex process governed by multiple factors.

\begin{figure}
	\centering
	\includegraphics[width=0.999\linewidth]{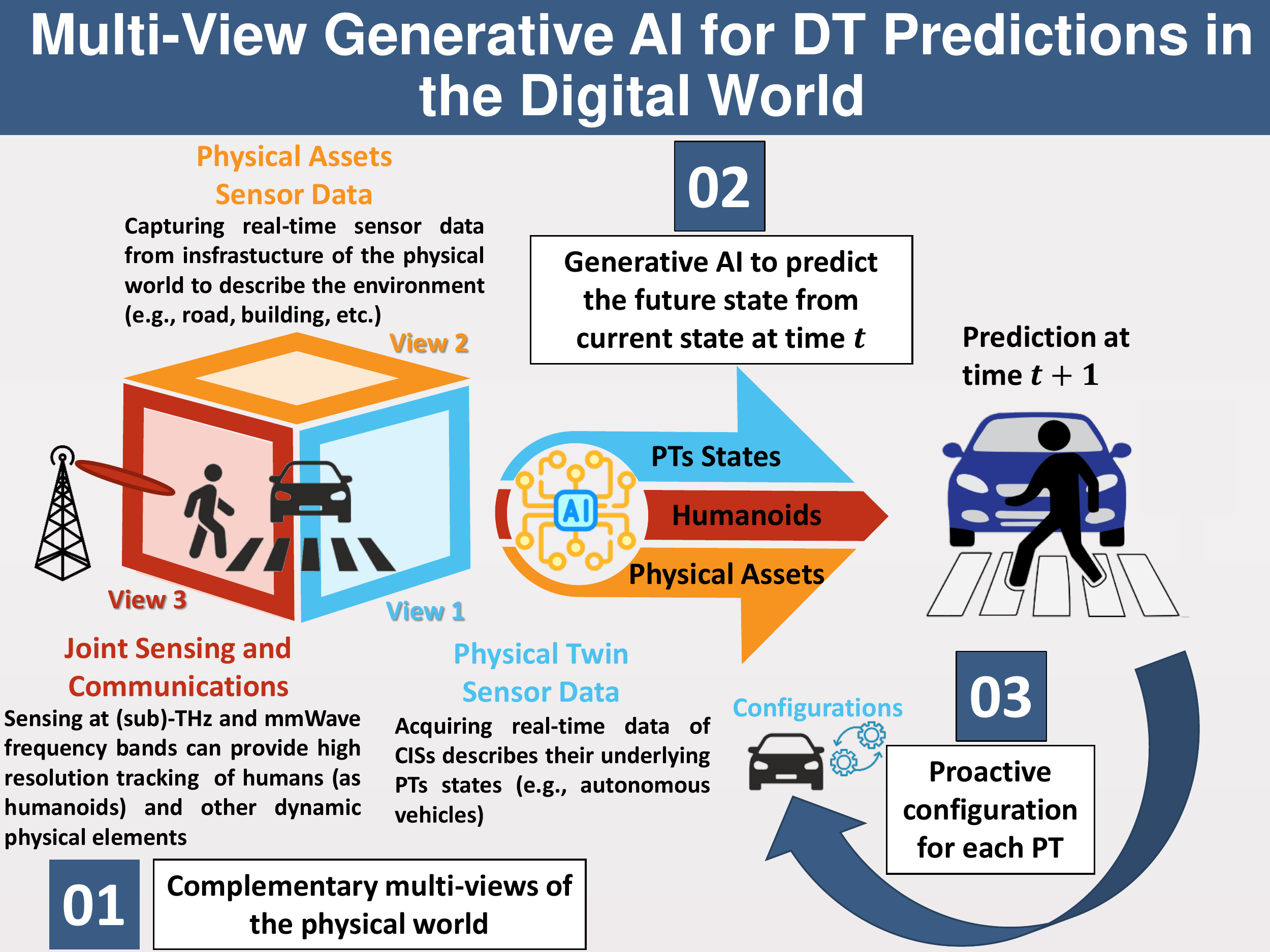}
	\caption{\small{Overview of the multi-view generative AI process for the DT configuration mechanism in the digital world.}}
	\label{Interactions}
 \vspace{-0.3cm}
\end{figure}

First, the future state of a \ac{CT} cannot be predicted \emph{individually} by relying just on the current state, as assumed in~\cite{lu2021adaptive}.
This is because \acp{PT} of \acp{CIS} are mutually \emph{correlated} in the real world, i.e., their future states could depend on their mutual, present states and actions. For instance, consider the example in Fig.~\ref{Interactions} with autonomous vehicles as \acp{CIS}. Each vehicle's future state depends on its current state as well as the actions and states of neighboring vehicles (e.g., a speed decrease by one vehicle impacts the speed of other vehicles).
Thus, any \ac{DT} predictive mechanism here should be capable of predicting the states of the \acp{CT} \emph{collectively}.
Second, as the \ac{CT} resides in the digital world, its environment will affect its state. Thus, the future state of a \ac{CT} must be predicted along with the future state of its digital environment.
This will implicitly require considering the state of the CT's counter physical environment in terms of \emph{physical assets} (e.g., road) in the prediction process. 
Third, \acp{CIS} will blend with humans in the physical world. Thus, the actions taken by the \acp{PT} are affected by the states of humans in their proximity. The future states of the \acp{CT} resulting from those will therefore depend on the \emph{situational state of the humans} (e.g., if humans are crossing the road, then the speed of the vehicles will be impacted in the future state).
Henceforth, to ensure adequate \ac{DT} configurations, it is pivotal to provide a novel prediction framework that integrates the: \emph{i)~states of \acp{PT}, ii)~states of the physical assets, and iii)~situational states of humans.}

One approach to tackle these challenges is to consider the aforementioned states as three viewpoints describing our physical system setting at time $t$. Then, these viewpoints are combined to predict the next setting using the framework of \emph{multi-view learning} -- a rigorous AI framework that allows the fusion of data generated from \emph{multiple views of the same subject}~\cite{xu2013survey}. In this framework, each data stream particularly describes the subject from one viewpoint, whereas their union provides a \emph{complementary} overview of the subject in hand. In our setting, the different viewpoints are analogous to the descriptions of the current states of \acp{PT}, physical assets, and humans in the physical setting at time $t$, as in Fig.~\ref{Interactions}. 
To address these ternary fronts, in the first view, the \acp{CIS} must utilize their collected \ac{PT} data to describe their states. 
In the second view, states of the physical assets could be captured through a distributed sensing architecture that uses the massive numbers of sensors in the physical environment~\cite{hashash2022towards}. 
In the third view, the network must capture the human presence via wireless sensing of their situational states. Thereby, their presence will be captured in the form of \emph{humanoids} in the digital world. To provide high resolution modeling of humanoids, joint sensing and communications at (sub-)\ac{THz} and \ac{mmWave} frequencies is a promising candidate~\cite{chaccour2022seven}. 

After acquiring multiple views of data at time $t$, the future system setting at $t+1$ (or generally additional states, i.e., $t+n$) is predicted by leveraging generative \ac{AI} abilities. This will implicitly provide the future states of the \acp{CT} within this physical environment. Finally, according to the predicted state of each \ac{CT}, the corresponding \ac{PT} is configured. This proactive configuration helps drive the \acp{PT} to their optimal performance and helps avoid any future setbacks.

\vspace{-0.3cm}
\section{Wireless, Computing, and \ac{AI} Challenges}
\label{Challenges}
In this section, we identify a plethora of wireless, computing, and \ac{AI} challenges that should be addressed to create a metaverse-ready wireless network that supports all metaverse worlds and experiences.
\vspace{-0.45cm}

\subsection{Metaverse Architecture}
To enable the cyber experience, we must deploy a synchronous digital world over the network. Nevertheless, replicating the real world in a centralized, cloud-based manner could incur high delays that can jeopardize this synchronization. To alleviate this challenge, a shift towards a \emph{decentralized, edge-enabled} digital world is indispensable.
Indeed, in~\cite{hashash2022towards}, we showed that the proper path to \emph{digital reality} is accomplished through decentralizing the digital world into \emph{sub-metaverses}, i.e., digital representations of spaces in the physical world. These sub-metaverses are orchestrated along with their constituents at the wireless edge.
In contrast, deploying the \emph{virtual world} is driven by the requirements of the extended, live, and parallel experiences.
On the one hand, the extended experience needs a centralized architecture for its applications (e.g., cloud gaming and social networking). On the other hand, the live and parallel experiences require the opening of a \emph{pipeline from the decentralized, digital world} into the virtual world. Therefore, a slice of the virtual world must be confounded with the digital world at the edge. In other words, the virtual world should be deployed in a \emph{semi-distributed} fashion, split between cloud and edge. 
Next, we delve into the challenges associated to the experiences and interactions within this architectural setting.
\vspace{-0.42cm}
\subsection{Wireless and Computing Challenges in the Metaverse}

\subsubsection{Preserving Synchronization and Homogeneity of the Digital World}
Conserving the synchronization between the real world and its decentralized sub-metaverses is perhaps the most pressing challenge for maintaining reliable predictions in the \ac{CIS}-to-\ac{DT} interaction as well as an immersive live experience. However, this decentralized metaverse needs precise orchestration to properly function.
Particularly, it is crucial to: \emph{1) synchronize each sub-metaverse with its counter physical space}, and \emph{2) preserve the homogeneity of the digital world} by keeping its different sub-metaverses in sync with one another. Hence, a critical challenge here is to minimize two types of synchronization: a) \emph{inter-synchronization time} between the physical and digital worlds to ensure a high fidelity replica of all physical assets and, b) \emph{intra-synchronization time} between the sub-metaverses themselves that is pivotal to conserve homogeneity. This requires new probabilistic or stochastic techniques that can effectively model and distribute the physical world under stringent wireless and computing resource budgets. 
We took a first step towards addressing this challenge in~\cite{hashash2022towards}. Therein, we showed how one can model  the physical world through a combination of spatial and sensing distributions.
Subsequently, we decomposed the physical world into sub-metaverses at the edge through an optimal transport theory technique that yields an optimal synchronization performance.

\subsubsection{Synchronization of Cognitive Avatars on the Network}
Since avatars are mainly \ac{AI} models, one wonders where the model of the cognitive avatar must reside. To maintain high synchronization with the \ac{XR} user, the avatar model should remain in its proximity. Thus, it is apropos to deploy the \emph{avatars at the edge} for computing purposes. Nevertheless, avatars interact in the virtual world (at the cloud or at another edge) within the extended, live, and parallel experiences. However, moving an avatar away from the edge to the desired destination (cloud or edge) will increase the synchronization mismatch with its corresponding \ac{XR} user. This can severely degrade the \ac{QoE}. 
Thus, enabling the avatar interactions without jeopardizing the synchronization and \ac{QoE} of the \ac{XR} user is a major challenge.
One can investigate the potential of \emph{semantic telepresence} to aid cognitive avatars, where the avatar can still reside at one edge and send its \emph{semantic clone} (or impact) to the other edge to interact with the elements there~\cite{chaccour2022less}. Here, the clone will return semantic feedback to the edge where the avatar is located, triggering a response in the human-to-avatar interaction.

\subsubsection{Ultra Synchronization, High Rate, Low Latency Communications for Human-to-Avatar Interactions} 
Existing wireless technologies that can deliver high rate and low latency for \ac{XR} users~\cite{chaccour2022seven} (e.g., \ac{THz} networks) cannot support our envisioned human-to-avatar interaction. In particular, while existing solutions can possibly guarantee a seamless \ac{XR} service in terms of HD frames and haptic feedback, they cannot sustain a synchronization of 1) senses and, 2) movements and actions between \ac{XR} users and their avatars.
Indeed, integrating \ac{XR} for avatars in the metaverse needs an additional dimension of \emph{ultra synchronization} over the network.
This is different from the previously discussed synchronization in that it is related to the \emph{delay gap} between the senses/actions themselves.
By achieving ultra-synchronization, the network can minimize the \emph{delay between the senses} which is not possible to guarantee with existing ultra reliable low latency communication approaches.
Moreover, ultra synchronization is necessary to guarantee the \ac{E2E} synchronization of movements and actions between \ac{XR} users and their avatars. This can be crucial for the real time fusion of actions in the \emph{avatar-to-avatar} interaction, which requires the precision of multi-dimensional actions (e.g., through arms, legs, etc.) to forth come together. Therefore, designing such ultra synchronization paradigm is major a network challenge.
\vspace{-0.3cm}
\subsection{\ac{AI} Challenges in the Metaverse}

\subsubsection{Resilience of \acp{DT} to de-synchronization}
It is evident that \ac{CIS}-to-\ac{DT} interactions occur in a \emph{non-stationary} open world setting. Thus, a \ac{CT} can encounter out-of-distribution data shifts with every change in the physical world of the \ac{PT}. This data shift can degrade the accuracy of the \ac{CT} model in twinning the \ac{PT} and eventually distort its predictions. To ensure the \emph{reliability} of this interaction, the \ac{CT} must adapt to the data shift by updating its underlying \ac{AI} model. This could yield a \emph{de-synchronization gap} between the twins~\cite{hashash2022edge}.
Since \acp{DT} must be \emph{hisory-aware}, this gap will continue to incrementally increase with each update phase. It is thus challenging to adapt the \ac{CT} model to reach utmost accuracy, while limiting this increasing gap.
Hence, the \emph{resilience} of \acp{DT} to de-synchronization will play a critical role in providing a swift return of the \acp{DT} into sync. To achieve such resilience, we need new \ac{AI} schemes to adaptively and rapidly update the \ac{DT} models to minimize the de-synchronization gap. Those \ac{AI} schemes must possess unique properties such as the ability to incorporate incremental knowledge. One promising approach is through the use of continual, lifelong learning~\cite{hashash2022edge}.
In addition, one can develop a semantic language between the \ac{PT} and \ac{CT} that allows the transmission of efficient representations describing the situation of the \ac{PT}.
If properly designed, this language can be robust to the variational data shifts, hence eliminating de-synchronization~\cite{chaccour2022less}. 

\subsubsection{Catastrophic vs. Graceful Forgetting for Generalizable \acp{DT}}
The key challenge to maintain a \emph{continuous} cyber experience is to sustain a \ac{CT} model that can \emph{generalize} over its history of physical world data. 
On the one hand, the \ac{CT} should be able to preserve \emph{the knowledge acquired} over its history. Hence, the \ac{CT} should \emph{mitigate catastrophic forgetting} of its knowledge that arises involuntarily while updating its \ac{AI} model and leads to a \emph{plastic} \ac{CT} model. Meanwhile, remembering the acquired knowledge \emph{persistently} will limit the \ac{AI} model's ability to adapt to new data streams, and will eventually yield a \emph{stable} \ac{CT} model. Therefore, it is pivotal, yet challenging, to balance the \emph{plasticity-stability dilemma} that should be addressed in the \ac{CT} model update process~\cite{parisi2019continual}. Additionally, to limit the influence of some random, infrequent instances in the \ac{PT} data, \emph{machine unlearning} can be leveraged to enable \emph{graceful forgetting} of such instances.


\subsubsection{Higher Order Reasoning for Cognitive Avatars}
One main challenge in the human-to-avatar interaction lies in the abilities of cognitive avatars to exploit the knowledge about their human persona, from the forward mode, to facilitate abductive reasoning, in the backward mode. To carry this out, the avatar should primarily determine the relationships between the user sensory input and their actions and movements. Thus, the effectiveness of abductive reasoning hinges on the clarity and consolidation of such relationships. To address this problem, avatars must be endowed with higher order reasoning capabilities. 
One possible approach is to investigate the \emph{relational reasoning} between the sensory inputs and actions to draw stronger conclusions. Such reasoning incurs a formal, higher order form of relationships, beyond statistical ones. This could involve the use of \emph{relational abstractions} for finding advanced formal conclusions~\cite{nam2022learning}.

\vspace{-0.3cm}
\section{Open Questions and Challenges}
Beyond the challenges of Section~\ref{Challenges}, we discuss several  open questions and challenges for the future generations of the metaverse:



    \subsubsection{Entropy of the metaverse} 
    The real world that the metaverse must replicate remains, at large, very random and unpredictable. Thus, it is necessary to provision a sufficient and significant amount of network resources to replicate this random structure. Here, to estimate the digital world resources, a fundamental metric worth investigating is the entropy of the metaverse that can provide a bound for the amount of resources needed under this randomness. This randomness also opens the doors for uncertainty analyses and how principles such chaos theory can apply to the metaverse.

    \subsubsection{Parallel metaverses}
    The coexistence of parallel metaverses can enable the omnipresence of \acp{CT} in different digital world settings. Here, one can ponder whether this omnipresence can account for different stochastic scenarios that \acp{CIS} can encounter in the physical world. This can pave the way for a more robust prediction scheme that deals with all possible consequences that can take place in the \ac{CIS}-to-\ac{DT} process.

    \subsubsection{Expandable metaverse}
    The portion of the metaverse that has been addressed to date is still infinitesimal in comparison to the extents of the real world. In fact, the metaverse could expand to engulf vast spaces in the world such as outer space, planets, and galaxies. For example, this can be pivotal for twinning the Internet of Space Things (e.g., satellites, planes, etc). Once the metaverse expands to this order, it will be challenging for the network to solely accommodate the digital world. Hence, the digital world will then extend beyond today's cellular network limits to be distributed over \acp{NTN} and mobile end computing devices such as satellites and drones.
    

    \subsubsection{Brain Controlled Avatars}
    Brain computer interfaces can enhance the \emph{virtual world experience} and aid conventional \ac{XR} sensing devices to control avatars. However, this will require envisaging new communication paradigms such as brain-to-brain communication and extracting the \emph{intent} of the \ac{XR} user from brain signals. Brain control allows the avatar to mimic the \ac{XR} user with higher fidelity that can surpass the abilities attained through \ac{AI}-based human reasoning.

\vspace{-0.25cm}
\section{Conclusion and Recommendations}
This paper presented a vision of a limitless, wireless metaverse while identifying its key constituents, worlds, experiences, interactions, and challenges. Building on the developed roadmap, we conclude with several recommendations:

\subsubsection{Towards a Digital World} Given that \ac{XR} and virtual world technologies are already underway, a first step toward a limitless metaverse should be to implement a complementary, scalable digital world.

\subsubsection{Advanced Immersive System} The metaverse will not be only an application of \ac{XR} and \acp{DT}. Instead, it represents an alternative parallel reality, driven by interactions between versatile constituents that should all be highly immersed to build this metaverse (see Fig.~\ref{Metaverse Vision}).
    
    \subsubsection{Synchronization -- a Key Challenge} Constituents such as avatars, \acp{CT}, and interactive assets impose new dimensions for synchronization. Achieving limitless metaverse synchronization requires sustaining continuous, real-time information pipelines between the worlds to facilitate seamless teleportation and interactions of the aforementioned constituents.

\subsubsection{Prominent Metaverse-Ready \ac{AI}} Realizing the metaverse vision requires deploying a new breed of AI algorithms with desirable properties such as reasoning, resilience, and generalization. Developing such metaverse-ready AI frameworks, while building on emerging tools (e.g., large language models), is necessary for a real world deployment of the metaverse.

\vspace{-0.3cm}


\bibliographystyle{IEEEtran}
\def\baselinestretch{0.96}
\bibliography{bibliography}
\end{document}

%% file: acronyms.tex
\acro{D2D}{device-to-device}
\acro{SIR}{signal-to-interference-ratio}
\acro{SINR}{signal-to-interference-plus-noise-ratio}
\acro{PCP}{Poisson cluster process}
\acro{CoMP}{coordinated multi-point}
\acro{BS}{base station} 
\acro{MD-CoMP}{macrodiversity CoMP transmission}
\acro{MAC}{medium-access-control}
\acro{JT-CoMP}{joint transmission CoMP}
\acro{CoMP-JT}{coordinated multipoint joint transmission}
\acro{SBS}{small base station}
\acro{MDSD}{multiple devices to a single device}
\acro{MDS}{maximum distance separable}
\acro{SCN}{small cell network}
\acro{PPP}{Poisson point process}
\acro{TCP}{Thomas cluster process}
\acro{CSI}{channel state information}
\acro{PDF}{probability distribution function}
\acro{PMF}{probability mass function}
\acro{RV}{random variable}
\acro{i.i.d.}{independently and identically distributed}
\acro{MBMS}{multimedia broadcasting multicasting service}
\acro{EE}{energy efficiency}
\acro{HCP}{hard-core placement}
\acro{CCDF}{complementary cumulative distribution function}
\acro{CDF}{cumulative distribution function}
\acro{PC}{probabilistic caching}
\acro{RC}{random caching}
\acro{CPF}{caching popular files} 
\acro{PGFL}{probability generating functional}
\acro{KKT}{Karush-Kuhn-Tucker}
\acro{PGF}{point generating function}
\acro{SCA}{successive convex approximation}
\acro{HD}{high-definition}
\acro{FHD}{full-high-definition}
\acro{UHD}{ultra-high-definition}
\acro{VR}{virtual reality}
\acro{AR}{augmented reality}
\acro{5G}{fifth-generation}
\acro{QoS}{quality-of-service}
\acro{QoE}{quality-of-experience}
\acro{IoT}{Internet of Things}
\acro{MHCPP}{Matern hardcore point process}
\acro{LoS}{line-of-sight}
\acro{NLoS}{non-line-of-sight}
\acro{PSD}{power spectral density}
\acro{MEC}{mobile edge computing}
\acro{E2E}{end-to-end}
\acro{THz}{terahertz}
\acro{CLT}{central limit theorem}
\acro{HQ}{High Quality}
\acro{eMBB}{enhanced mobile broadband}
\acro{URLLC}{ultra reliable low latency communications}
\acro{mmWave}{millimeter wave}
\acro{EVT}{extreme value theory}
\acro{GEV}{generalized extreme value}
\acro{LIS}{large intelligent surface}
\acro{RIS}{reconfigurable intelligent surface}
\acro{RF}{radio frequency}
\acro{UE}{user equipment}
\acro{MIMO}{multiple-input multiple-output}
\acro{EVaR}{entropic value-at-risk}
\acro{DNN}{deep neural network}
\acro{MDP}{Markov decision process}
\acro{RL}{reinforcement learning}
\acro{RNN}{recurrent neural network}
\acro{ANN}{artificial neural networks}
\acro{LSTM}{long short-term memory}
\acro{ReLu}{rectified linear unit}
\acro{VaR}{value-at-risk}
\acro{SNR}{signal-to-noise ratio}
\acro{AoSA}{array of subarray}
\acro{XR}{extended reality}
\acro{AoA}{angle of arrival}
\acro{ULA}{uniform linear array}
\acro{AoD}{angle of departure}
\acro{EM}{electromagnetic}
\acro{HRLLC}{s high-rate and high-reliability low latency communications}
\acro{6DoF}{six degrees of freedom}
\acro{MR}{mixed reality}
\acro{PAPR}{peak to average power ratio}
\acro{OFDM}{orthogonal frequency-division multiplexing}
\acro{OFDMA}{orthogonal frequency-division multiple access}
\acro{SC-FDM}{single carrier frequency-division multiplexing}
\acro{ToA}{time of arrival}
\acro{MUSIC}{multiple signal classification}
\acro{IoE}{Internet of Everything}
\acro{DT}{digital twin}
\acro{PT}{physical twin}
\acro{CT}{cyber twin}
\acro{DRL}{deep reinforcement learning}
\acro{FL}{federated learning}
\acro{DL}{deep learning}
\acro{CRAS}{connected robotics and autonomous system}
\acro{CL}{continual learning}
\acro{MSE}{mean squared error}
\acro{EWC}{elastic weight consolidation}
\acro{ML}{machine learning}
\acro{GD}{gradient descent}
\acro{MLP}{multi layer perceptron}
\acro{TL}{transfer learning}
\acro{AI}{artificial intelligence}
\acro{NFT}{non fungible token}
\acro{H2A}{human-to-avatar}
\acro{A2A}{avatar-to-avatar}
\acro{UAV}{unmanned aerial vehicle}
\acro{NTN}{non-terrestrial networks}
\acro{CIS}{connected intelligence system}
\acro{QoVE}{quality of virtual experience}
\acro{OOD}{out-of-distribution}